\documentstyle[twoside, epsfig]{article}
\input pz.sty

\begin{document}
\PZhead{5}{31}{2011}{October 1}{October 17}

\PZtitletl{Photometric Mass Estimate for the Compact Component of SS 433.}
{And Yet It Is a Neutron Star}

\PZauth{V.~P. Goranskij}

\PZinsto{Sternberg Astronomical Institute, Moscow University,
Universitetski pr., 13, Moscow, 199992, Russia; e-mail: goray@sai.msu.ru}

\SIMBADobj{V1343 Aql}

\PZabstract{After 33 years of extensive studies of SS 433, we have learnt
much about this unique system with moving emission lines in the spectrum.
The orbital inclination is known from spectroscopic observations of moving lines;
the distance is derived from radio interferometry of relativistic jets;
the mass ratio of its components is determined from X-ray observations
of jets' eclipses. In 2005, the accretion donor was
detected as an A4 $-$ A8 giant, and its contribution to eclipse light was
measured spectroscopically. In the present paper, the A-type star was detected
via multicolor photometry on the basis of its Balmer jump.
A method is proposed to estimate the interstellar reddening, able
to measure the individual law of interstellar absorption for SS 433 from
spectrophotometry. The method is based on the extracting the energy distribution
of the spectral component of a very hot source covered in eclipse
and on the comparison of its energy distribution to the Planck energy
distribution of a black body with the temperature exceeding 10$^6$ K.
The determination of general parameters of SS 433 leads to fairly accurate
estimates of luminosity, radius and mass of the A star in the system,
and consequently leads to an accurate estimate of the mass of the compact component,
the source of jets. This latter mass is between 1.25 and 1.87 solar masses.
The reasons of overestimating this mass when using the dynamical method
are discussed. In our opinion, the presence of a black hole in this system
is excluded.}

\section{INTRODUCTION}

SS 433 (Stephenson \& Sanduleak, 1977) = V1343 Aql is a unique variable star
with moving emission lines in the spectrum.
Each stationary emission of Balmer or \hbox{He I} lines has two
moving emission components that are formed by a pair of oppositely directed
and highly collimated precessing relativistic jets or beams of very hot
matter accelerated to the velocity of 0.26{\it c}, i.e. to ~80000 km/s.
The period of jet precession is 162$^d$.5 $-$ 164$^d$
(Margon et al., 1979; Margon, 1984).
The jets are inclined to the precession axis by 19\fdeg75, and the
axis of precession is inclined by 78\fdeg81 to the line of sight
(Davydov et al., 2008). Like wavelengths of other moving lines, the
wavelenghts of H$_\alpha$ components follow sine functions with
the full amplitudes of 1200 \AA\, in antiphase (Fig.~1). In
this motion, H$_\alpha$ components are superimposed in the position located
near 6800 \AA\ twice for a precession period, at the times usually designated
T$_1$ and T$_2$. The wavelength where the components coincide differs from
the stationary H$_\alpha$-line wavelength by 250 \AA, and this is due to
transverse Doppler effect following from the special theory of relativity.
The time of largest divergence of moving emissions is designated T$_3$.
Additionally, the jets show nodding oscillations
(or jitter) with the period of 6$^d$.28  (Newsom \& Collins II, 1981 and 1986;
Wagner et al., 1981) and amplitude of 2\fdeg8 (Borisov \& Fabrika, 1987).
The matter of jets is erupted into "bullets" that become apparent as
discrete line components in the spectra,
so that each emission detail appears at a fixed wavelength, it strengthens
in a typical time range of about 0.5 $-$ 1$^d$  and decays in 1 $- 3^d$
(Grandi \& Stone, 1982; Vermeulen et al., 1993a). Several bullets form the
composite structure of a moving emission line.
The collimation angle of a bullet is about 1\fdeg0 - 1\fdeg4
(Borisov \& Fabrika, 1987).

The detales of precessing jets may be followed by radio interferometry in
the milliarcsecond scale (Vermeulen et al., 1993b). Deep radio
images were taken with the VLA radio telescope, they show the structure of jets
up to angular distances of 4\arcs (Blundell \& Bowler, 2004). Fitting radio
images with the kinematic model of jets gives a sufficiently accurate
distance to SS 433, independent on interstellar absorption. This distance is
in the range between 4.85 (Vermeulen et al., 1993b using the VLBI) and
5.50 kpc (Hjellming \& Johnson, 1982, from the VLA, and Blundell \& Bowler, 2004,
from the VLA). Intermediate distance values
were measured with the same method by Spencer (1984),
MERILIN, 4.9 kpc; Fejes (1986), VLBI, 5.0 kpc; and Romney et al. (1987),
VLBI, 5.0 kpc. There is no systematical trend of distance measures versus time,
and the mean distance value is 5.12 $\pm$0.27 kpc.

\PZfig{11cm}{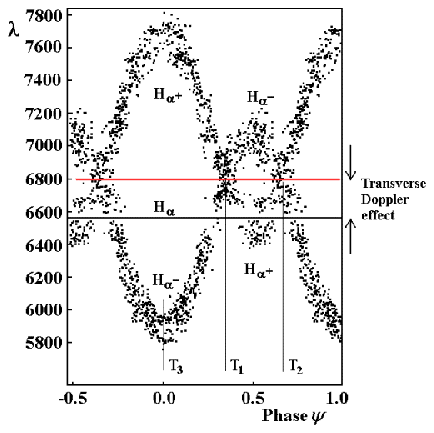}{Wavelengths of the H$^+_\alpha$ and H$^-_\alpha$
moving components versus precession phase $\psi$. The black horizontal line
marks the location of the stationary H$_\alpha$ emission line. The red horizontal line
is the axis of symmetry passing through the points of coincidence of the moving
components T$_1$ and T$_2$. The zero point of precession phase is adopted at
T$_3$.
}

SS 433 is also known also an eclipsing binary system with the orbital period of
13$^d$.082 (Gladyshev et al., 1980; Gladyshev, 1981; Cherepashchuk, 1981;
Crampton \& Hutchings, 1981). In kinematical jet-precession models derived
from observations of the moving lines, the axis of
precession coincides with the orbital axis, so the orbital inclination of SS 433
is measured correctly. Photometry of SS 433 reveals precession and nodding
periods, as well as light outbursts at different time scales (Goranskii et al.,
1998a). The shape of eclipsing light curve and its maximum light depend on the
phase of precession period (Fig.~2). Near T$_3$ times, it looks like that of
a typical $\beta$ Lyrae variable star having two minima of different depths.
At other phases of precession, the dispersion of observations increases, but
some orbital-phase averaged light curves resemble those of Cepheids. There are
pronounced brightness variations in eclipse depth depending on the
precession-period phase $\psi$,
with the $V$-band amplitude of 0\fmm4.
These variations suggest that the eclipses are partial.
However, the same variations can be fitted by
the nodding period as well because the precession and nodding periods are
related to the orbital one by the formula:\\

$1/P_{nod} = 2/P_{orb} + 1/P_{prec}$, \ \ \ \ \ \ \ (1)\\

\noindent
and thus, phases of nodding and precession periods are similar if the observations
are chosen from a narrow range of the orbital period (during the eclipse).
This means that we cannot distinguish precessing and nodding light variations
in the eclipse depth.

\PZfig{4.7cm}{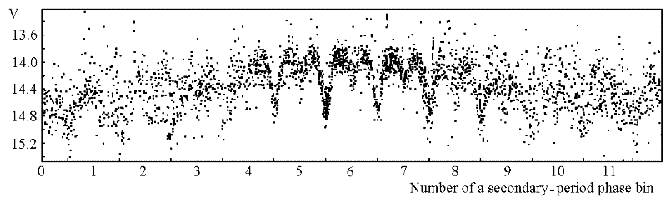}{Evolution of the orbital light curve
with the phase of the secondary precession period. To demonstrate the effect,
the orbital-period light curves are calculated in 12 intervals (bins) of the
precession period, and displayed in the order of increasing precession phase.
The precession zero phase $T_3$ corresponds to the bin No.~6. Several observations are
out of this magnitude range and are not plotted.
}

Multicolor photometry reveals peculiar behavior of SS 433 in different filters.
If the $U-B$ and $B-V$ color indices remain approximately constant and independent
on brightness, the $V-R$ index varies with the orbital phase, being strongly
correlated with the $V$ or $R$ magnitudes. During flares or active states,
the $V-R$ index departs from this relation to the right side (Gladyshev, 1981),
suggesting that the deviations are connected with a brightening of the
stationary H$_\alpha$ emission.
However, observations in the $UBVRI(H_\alpha)$ filters in a flare performed by
V.~Rakhimov (Aslanov et al., 1993) revealed that the amplitude of the flare
in the near-infrared $I$ filter ($\lambda_{eff}$ = 7700 \AA; $FWHM$ = 1040 \AA)
was larger than in the narrow-band $H_\alpha$ filter ($\lambda_{eff}$ = 6550 \AA;
FWHM = 200 \AA). This phenomenon was noticed by an anonymous referee of the
{\it Astronomy and Astrophysics}, so we call it "the anonymous referee effect".
The referee's finding led to a chain of further discoveries.
First, it was found that the light curve of SS 433 consists of two almost
independent light curves formed by two light sources having essentially
different $V-R$ color indices, 1\fmm9 and 2\fmm8. Second,
both light curves may be easily extracted from the combined light curve. Third,
the blue source shows periodic orbital, precessing and nodding variations,
whereas the red one is not periodic and exhibits flares that
coincide with radio flares (Goranskij et al., 1998b).
It became clear that the spectrum of red source
extended up to radio wavelengths. Its contribution to the combined
spectrum of the system increased to longer wavelengths. The red source
veils the blue source, and therefore amplitudes of the orbital and precession
light variations decrease from $U,B,V$
to $J,H,K$ bands. The contribution of the blue source is variable with the
orbital period, and this is a reason why the $V-R$ index varies with the orbital
phase.

The mean $U - B$ and $B - V$ colors of SS 433 are respectively 0\fmm90 and 2\fmm20.
They are strongly subjected to interstellar reddening. These indices suggest
that SS 433 is a hot OB star with the reddening $E(B-V) \approx$ 2\fmm53
and the full extinction
$A_V$ of approximately 8\mm\ (Murdin et al., 1980; Cherepashchuk et al., 1982;
Margon, 1984). Wagner (1986) found $A_V$ = 7.8\mm\ $\pm$0.5\mm\ from
continuum fitting of spectrophotometric observations in the wavelength range
between 4036 and 8100 \AA. The mean color temperature was about 32500 K,
the continuum spectrum appeared hotter when precessing body was brighter, and
these variations corresponded to an amplitude of 0\fmm08 in the $B - V$ color.

The most valuable information on jet structure and mass ratio of the components
of SS 433 can be derived from X-ray observations of its eclipses. The Ginga X-ray
observatory has made the most important contribution to the research of SS 433.
Three eclipses were continuously monitored with a complete phase coverage at
different phases of the precession period, namely those in May 1987, May 1988 and
May 1999; additional pointings to the object were performed at different
orbital and precession phases (Yuan et al., 1995). It became clear that
most of X-rays were emitted from jets. The shape of the continuum could be fitted
with a single thermal bremsstrahlung model, its characteristic temperature
$kT_0$ being
about 20 keV. With the eV-to-Kelvin reduction coefficient of 1.1604$\cdot10^4$,
this temperature is 232 $\cdot 10^6$ K. Thus, there exists an uncovered source with
a temperature that high in the system. Note that one cannot distinguish thermal
sources with temperatures above 10$^6$ K using multicolor
$UBVRI$ photometry because their spectral energy distributions (SEDs) have
maxima far in the X rays and their slopes are similar in the $UBVRI$ spectral
region. Such energy distributions do not differ much from those of
an O-type star with a temperature about 32500 K (Wagner, 1986), and their
color temperature can be underestimated due to the contribution from
the red and infrared excess, discovered with the help of the referee.
Ginga observations also show a strong K emission blend of Fe~XXV/Fe~XXVI at
6 keV that can be extracted from the spectrum. The line shows the same
relativistic shifts as the moving "violet" (short-wavelength) components
of the Balmer emission. Some observations made near T$_1$ and T$_2$
also display a moving "red" component from the oppositely directed jet.

X-ray eclipses of SS 433 are partial and show strongly pronounced contacts
of the jet base with the companion's limb. One of the eclipses registered by
Ginga, centered at 1987 May 20.9 $\pm$0.5 UT (precession phase $\psi = 0.25$
if counted from T$_3$),
occurred in the high state (a strong flux and hard spectrum) (Kawai et al., 1989;
Brinkmann et al., 1989). The duration of the eclipse measured as the time
interval between contacts is 2$^d$.4. The shape of the eclipse is similar
in different energy bands, the intensity of the extracted K$\alpha$ line shows
variations with the same contacts and the same relative depth.
This observation confirms that it is the jet that is eclipsed, and the
jet is the brightest X-ray source in the system.
Additionally, these observations along with
the well-known orbit inclination enable us to measure the mass ratio of the components
very accurately. With the standard assumption that the optical star fills its
Roche lobe and using Paczynski's formula for the size of the Roche lobe,\\

$R_*/a = 0.38 -0.2 \cdot log_{10}(q) \approx 0.545$ \ \ \ \ \ \ \ (2)\\

\noindent
in this case, Brinkmann et al. (1989) obtained the ratio $q = M_x/M_{opt} = 0.1496$.
In Paczynski's formula, $R_*$ is the radius of the optical star and $a$ is the
binary's separation. Using the results of Doppler analysis of the stationary
He II lines by Crampton and Hutchings (1981),
Brinkman et al. (1989) derived the masses 2.1 $M_\odot$
and 14 $M_\odot$ for the components and the optical component's radius
32 $R_\odot$. With such a mass ratio, one should expect that, in eclipses,
the donor covers the full Roche-lobe volume of the compact companion, including
the surrounding accretion disk if it exists. This simple solution is based on
the assumption that the jet is thin, i.e. its width is negligibly small compared
to the radius of the optical star.

Kawai et al. (1989) considered solutions for $q$ based on more wider assumptions
like opposed jets, a thick accretion disk around the compact object, a thick jet
with a cylindrical axisymmetric shape. These assumptions disperse the values of $q$
in the 0.13 - 4.6 range. Different authors considered a cocoon
instead of a jet, or an eclipse of the jet by a star having an extended outflowing
envelope without any clear-cut limb. All these attempts were made to increase
$q$ ratio. In my opinion, a thin jet is a satisfactory solution. Actually, no
additional contacts are seen in the shape of X-ray eclipses that would
look like intensity jumps. In the case of a thin jet, the exponential
intensity decrease along its length can be revealed from observations, it is described
with the following formula:\\

$I(x) = a \cdot exp(-1.8x/R_*)$, \ \ \ \ \ (3)\\

\noindent where $x$ is the length, measured from the base.
This formula can be derived with the technque of lost light as follows. First,
we interpolate the out-of-eclipse flux to the phase range of the eclipse and measure the flux
deficiency due to the eclipse for each in-eclipse observation. Second, we calculate
the covered length of the jet for the phase of each point in a simple model.
The relation of flux deficiency versus covered
length can well be fitted with an exponent function; the points of ingress
and egress fall on the same line in spite of the different rate of variation.
A steep ingress of this eclipse means that the base of the thin jet is directed
along the star limb, and the base disappears rapidly. Contrary, the egress is slow;
this means that the jet base reappears from behind the limb at a large angle. This behavior
corresponds to precession phase 0.25 for the retrograde motion of the jet with respect
of the orbital motion. The relation of intensity versus jet length can be
revealed by differentiating the averaged flux deficiency function.
Naturally, we would not be able to see such a behavior were the jet thick, and were the
thickness of the jet covered in ingress or egress.
Our model light curves for May 1987 X-ray eclipse were published by
Goranskij et al. (1997, Fig.~2), and by Goranskij (1998c, Fig.~2).
From this model, with the jet base located in the orbital plane, we found
$q = 0.155 \pm0.015$. This value may be even less if the base of the visible jet is
above the plane. This is in a very good agreement with the results from Brinkmann's et al. (1989).

ASCA high-resolution X-ray spectral observations performed on
April 23 and 24, 1993 resolved the K$\alpha$ blend at 6 keV into
Fe~XXV and Fe~XXVI lines; each line showed both a strong
"violet" component and a faint "red" one (Kotani et al., 1994).
Both components had large Doppler shifts,
and positions of components in the spectra agreed with the kinematical
model. Detections of moving lines of Ar~XVII, Ar~XVIII, S~XV, S~XVI, Ca~XIX, and
Ni~XXVII ions were also reported. XMM observations (Brinkmann et al., 2005)
also confirm large Doppler-shifted emission components; the "red" jet contributes
30 - 40 percents to the total photon flux. The best estimates of the temperature
at the jet base are around $kT_0 \sim 17 \pm2$ keV. The "stationary line"
in K$\alpha$ around 7 keV was found in the Ginga data, but not in later ASCA
and Chandra observations (Brinkmann et al., 2005). The XMM K$\alpha$ line profiles
seem to be strongly affected by the "stationary line", which may be an indicator
of Compton scattering of jet base radiation on surrounding cold matter.
One eclipse, dated by 2003 May 11, was observed by the INTEGRAL X-ray observatory
in hard X-rays ranged between 25 and 100 keV during a large campaign
including optical spectroscopy, optical and infrared photometry
(Cherepashchuk et al., 2005). Unfortunately, the
phase covering of the eclipse in hard rays was insufficient.
Nevertheless, fragments of light curves suggest that the width of the hard X-ray eclipse
is larger than that in soft X-rays; probably the width increases with energy. The
hard X-ray eclipse is at least two times deeper then the soft X-ray eclipse.
No contacts were seen. Thus, the authors treat the eclipse as an occultation not of
a jet but of a "corona" surrounding the accretion disk.

Finally, weak lines of an A-type star, a probable mass donor were discovered
during eclipses in the spectra of SS 433 (Gies et al., 2002a).
These lines are present in the blue spectral range. The lines of the A star
are so weak, its spectroscopic contribution in the mid-eclipse
is estimated as 0.36 $\pm$ 0.07 \% in the V band (Hillwig \& Gies, 2008).
After this discovery, several attempts were made
to determine masses of SS 433 components with the dynamical method, i.e.
measuring amplitudes of the radial velocity curves, $K_x$ and $K_{opt}$.
The results are reviewed in Kubota et al. (2010).
They are the following.

\begin{center}
\begin{tabular}{lccc}
Source                     & $M_x$        & $M_{opt}$     & $q$  \\
\hline
Gies et al. (2002a)        &  11 $\pm$5   & 19 $\pm$7    & 0.58 \\
Gies et al. (2002b)        &  16.6 $\pm$6 & 23 $\pm$8    & 0.72 \\
Hillwig et al. (2004)      &  2.9 $\pm$0.7& 10.9 $\pm$3.1 & 0.27 \\
Cherepashchuk et al. (2005)&  18          & 24            & 0.75 \\
Hillwig \& Gies (2008)     &  4.3 $\pm$8  & 12.3 $\pm$3.3 & 0.35 \\
Kubota et al. (2010)       &  1.9 $-$ 4.9   & 10.4$^{+2.3}_{-1.9}$& 0.18 - 0.47\\
\end{tabular}
\end{center}

However, Barnes et al. (2006) argued that the A-type supergiant spectrum
may not be formed in the photosphere of the donor, it probably originates in the
accretion disk wind. In such a case, the visibility of absorption lines
should decrease in eclipse, but the reverse effect is observed. This is a good
reason to believe that an A-type donor exists in the system.

Whatever the source, the donor or the wind, the data ambiguity
concerning the components' masses reflects difficulties of the dynamic method.
These difficulties are the following. With $q$ as small as 0.15 and with
a very hot radiation source heating the surface of the A-type star,
we can observe the absorption-line spectrum only from the opposite side of
the donor. The mass center of the system is located near the center of the donor, so
actually we see the effect of donor's synchronous rotation, not the orbital
motion. This effect leads to an overestimate of the amplitude of mass-center motion.
Kubota et al. (2010) tried to account for this effect, and the lower limit of the
mass estimate decreased to 1.9 $M_\odot$ as a result. With a simple
model, they reduced $K_{opt}$ to 40 $\pm$5 km s$^{-1}$. They conclude that
"the compact object in SS 433 is most likely a low mass black hole, although
the possibility of a massive neutron star cannot be firmly excluded".
Otherwise, the mass of a neutron star may be close to the Chandrasekhar limit of
1.4 $M_\odot$ if SS 433 is a semidetached system with a heavy accretion disk
(having a mass of 0.5 $M_\odot$), or, what is more probable, a contact system
with an 0.5 $M_\odot$ star-like component having a neutron core
[the Thorne \& Zhitkov (1975, 1977) case]. Certainly, one should take
into account the presence of other matter in addition to the compact star in its
Roche lobe in the system with a large a rate of mass exchange.

Additionally, most authors used $K_{opt}$ = 175 km s$^{-1}$,
as measured by Fabrika \& Bychkova (1990) on the base of the He~II 4686 \AA\ line
or $K_{opt}$ = 162 km s$^{-1}$ from Gies et al. (2002b), based on the C~II 7231,
7236 \AA\ blend. These emissions were attributed by the cited authors to the accretion
disk surrounding the compact companion. This is a weak assumption. Typically,
the He~II line has $FWHM$ = 700 km s$^{-1}$, with wings to $FWZI >$ 2600 km s$^{-1}$.
In an eclipse, the equivalent width increases due to fainter continuum, so
the largest part of the emission belongs to a nebular envelope surrounding the
system but not to the accretion disk. Indeed, its shape is sometimes double-peaked.
Howerever, in an eclipse, we should expect, first, covering the blue peak from
the approaching disk stream; then, covering the red peak from the recessing stream;
and then, their consequent recovering. Actually, the eclipse in He~II line
goes otherwise. The blue wing of the line is eclipsed, and the line becomes
narrower, what Goranskij et al. (1997) treat as an eclipse of a wide-angle conical
outflow directed along the jet. Our analysis of the spectra by Kubota et al.
(2010) taken in the eclipse on 2007 October 6 confirms this behavior
of the He~II line. Thus, problems of dynamical mass determination for SS 433 are
due to complex structure of lines that do not reflect the mass-center motion of
the components.

Alternatively, we can find masses of the components in the system of SS 433
from the spectroscopic contribution of the A-type donor during an eclipse (36 $\pm$7
percent in the $V$ band). For the
known upper limits of distance ($d \le$ 5500 pc), and interstellar extinction
($A_V \le$ 8\fmm3) we can calculate the lower limit of the donor's absolute
magnitude, $M_V$ = $-6$\fmm2 from a simple formula:\\

$M_V = m_V + 5 -5 \cdot log_{10}(d) - A_V$. \ \ \ \ \ \ \ \ (4)\\

\noindent
Based on the theoretical and observational mass-luminosity
relations, such a star may have mass of 11.7 $M_\odot$. Then, the upper
limit for the compact companion is 1.76 $M_\odot$ taking into account
$q$ = 0.15, as derived from X-ray observations. Thus, the modern knowledge of
SS 433 parameters currently seems to exclude the presence of a black hole in SS 433.

Nevertheless, I tried to verify and improve the modern data on mass ratio,
interstellar extinction, photometric calibration using contemporary optical
and X-ray observations. It is also of interest to estimate the contribution
from the A-type donor using the photometric method.

\section{ORBITAL PERIOD AND MASS RATIO}

To check the mass ratio determined with GINGA X-ray observatory, I used 79784
X-ray observations of RXTE/ASM in the 1.3 $-$ 15 keV energy range accumulated
between 1996 January 5 and 2011 March 9 at the RXTE
Internet site:
\begin{verbatim}http://xte.mit.edu/ASM_lc.html.\end{verbatim}
These observations have a very low accuracy,
but with the number of measurements that large, if averaged in orbital or precession
phase bins, they allow us to establish the shapes of precession and eclipse light
curves in this energy range. The orbital period was improved using our collection
of photometric UBVRI observations. The light curves and literature source
list can be viewed with a Java-compatible browser at:
\begin{verbatim}http://jet.sao.ru/~goray/v1343aql.htm.\end{verbatim}
The table of observations is accessible in the file:
\begin{verbatim}http://jet.sao.ru/~goray/SS.DAT.\end{verbatim}
The first column of the Table
contains Julian Dates of observations in a form of JD-2400000.0; the next columns
are respectively  the $V, B, U, R$, and $I$ magnitudes, with their source
indicated in the last column. Note that the $R$ and $I$ magnitudes were taken both in Johnson and
in Cousins systems, so the light levels are highly inconsistent. These observations
were performed in the time range between 1978 October 5 and 2011 July 29,
and the Table is continuously replenished by new data.

The 35 best revised and new mid-eclipse times are given in Table 1.
To determine these times, the eclipse light curves were superposed with the same,
but mirrored, curves in the PC screen, the axis of the mirror reflection was
taken for the center of an eclipse. With these eclipse timings, the light ephemeris
was determined using least squires fitting, with the following result:\\

\begin{tabular}{crrl}
$Min I = $&$JD hel.$ 2450023.746 & + $13^d.08223$&$\cdot$E. \ \ \ \ \ \ \ (5)\\
         &            $\pm$.030 &   $\pm$.00007&\\
\end{tabular}

\vspace{2mm}
\noindent
The O$-$C curve calculated with this ephemeris is shown in Fig.~3 (top). Note that
the deviations of individual minima from the ephemeris are larger than their
mean errors. Fig.~3 (bottom) demonstrates that these deviations do not depend
on the precession phase. Precession phases $\psi$ were calculated
using the following ephemeris:\\

$T_3 = JD\ 2449998.0 + 162^d.278\ \cdot\ $E. \ \ \ \ \ \ \ \ (6)\\

Fitting the O$-$C curve with a quadratic expression reveals the quadratic term
to be insignificant. This means that the orbital period did not vary between 1979
and 2007.

\PZfig{11cm}{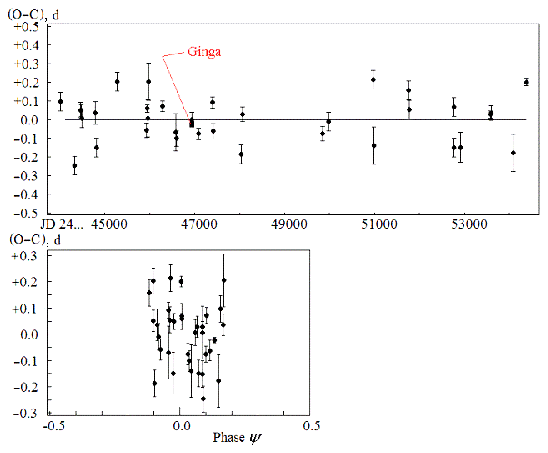}{The O$-$C curve for the orbital period (top).
Deviations of eclipse minima plotted versus phase of the precession period (bottom).
No relation between them is visible.
}

The light curve for RXTE/ASM data in the 1.3 $-$ 15 keV energy range
plotted versus phases $\phi$ of the orbital period calculated from
eq. (5) is shown in Fig.~4 (top). The points deviating from zero
by more than $\pm$10 cts s$^{-1}$ were eliminated. Fig.~4 (middle) demonstrates
the same data
plotted versus phases $\psi$ of the precession period calculated with
the formula (6). This is a mean light curve averaged by phase using the moving-
average method, with a phase interval of 0.10. The curve indicates periodic
variability of the X-ray flux in the range between 0.15 and 0.55 cts s$^{-1}$,
i.e. by the factor of 3.7. Fig.~4 (bottom) demonstrates the residuals revealed by
prewhitening the data for the precession wave, plotted versus the
orbital-period phase again. The residuals were averaged by phase $\phi$
calculated from eq. (5)
using the same moving-average method but with a smaller phase interval of 0.05.
The mean orbital light curve shows the average flux constant in the phase range
between 0.10 and 0.90, with a dip at phases between 0.90 and 0.10 due
to the eclipse. The RXTE/ASM light curve is compared to the GINGA light curve
for the eclipse on 1987 May 20 in Fig.~5.
Taking into account a small systematic shift of
the individual eclipse relative to averaged curve of multiple eclipses and
smoothing the averaged curve with the width of average interval,
one finds the agreement to be enough. RXTE/ASM data confirm that the
eclipse is caused by covering the jet base, and the jet base is located as close
to the donor's surface as follows from the mass ratio of $q$ = 0.15.

\PZfig{11cm}{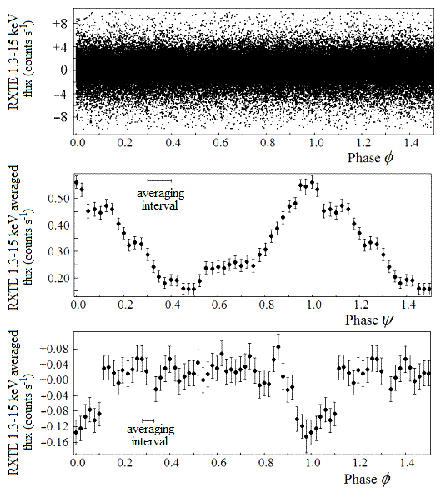}{The RXTE/ASM X-ray phased light curves.
Top: 79784 X-ray observations plotted versus the phase of the orbital
13$^d$.08223 period. Middle: the averaged light curve calculated with the precession
period of 162$^d$.278. Bottom: the averaged light curve for residuals after
prewhitening the precession wave plotted versus the phase of the orbital period,
13$^d$.08223. The middle and bottom curves were calculated using the moving-average
method, the averaging intervals are shown.}

\PZfig{11cm}{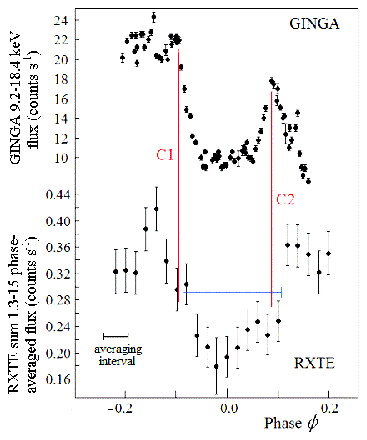}{Comparison of X-ray eclipse light curve shapes.
Top: the GINGA light curve for the eclipse on 1987 May 20. Bottom: the average eclipse
light curve from RXTE/ASM data. C1 and C2 are phases of jet-base contacts with
the donor's limb.}

\section{THE INTERSTELLAR EXTINCTION LAW}

SS 433 is subject to strong  extinction by interstellar medium. This extinction
is a  sum of absorption and scattering of the star light by the
interstellar dust; physics of these processes is described by
Savage \& Mathys (1979). The $V$-band extinction of SS 433 light may reach
8 magnitudes. To estimate the absolute magnitude of the A-type donor of SS 433,
we need to know both the accurate reddening $E(B-V)$ and the shape of
the interstellar-extinction curve, $A_1(\lambda)$, normalized to  unit
reddening $E(B - V)$ = 1.0. The extinction curve was a subject of many
studies; cf. the historical review in Straizys (1977). It was
ascertained that the shape of the extinction curve varied in different
directions and had different anomalies both in the near-ultraviolet and
near-infrared ranges. Strong anomalies were found in the
ultraviolet range in the nearby
LMC and SMC galaxies. In Fig.~6, we compare the interstellar extinction law curves
measured by J. Sudzius (published in Strayzys, 1977) for the directions of Cepheus,
Perseus, and Monoceros (converted from optical depths to magnitudes) to
later-published curves by Schild (1977), Savage \& Mathys (1979), Seaton
(1979), Koorneef \& Code (1981), Nandy et al. (1981), Howarth (1983), and
Prevot et al. (1984). Most authors agree that $A_V$ = 3.08 at $E(B - V)$ = 1.0,
but this value may be somewhat variable.

The formula to calculate interstellar extinction $A(\lambda)$ is:\\

$A(\lambda) = A_1(\lambda) \cdot E(B-V)$. \ \ \ \ \ \ (7)\\

Observations show some deviations in the $U$-band extinction law that
can reach 0\fmm8. The largest deviation downward found in the ultraviolet is
exhibited by the extinction curves in the LMC and SMC by Nandy et al. (1989), curve 5;
Koornneef \& Code (1981), curve 6; and Howarth (1983), curve 8a.
Curve 8 by Howarth (1983), measured for the Galaxy, does not show a deviation that big.
The largest deviation downward for the Galaxy is in the curve 6 by
Schild (1977). This author found anomalous absorption below the Balmer jump
in Perseus and in one region in Cygnus, but the normal law in the other Cygnus
region. According to Galactic data, the dispersion of extinction curves
in the $UV$ band in our Galaxy may reach 0\fmm5.

I have chosen the Savage \& Mathys (1979) curve as a template to use for SS 433
because it is in the best agreement with other observations in $BVR$ spectral range
and is frequently used in data reduction with ESO MIDAS package. It should be taken
into account that this table of extinction can have an insufficient accuracy when
applied to reductions for a star as absorbed as SS 433.

Fortunately, there is a method to correct the individual extinction law for SS 433
on the base of multicolor photometry or spectrophotometry. Using observations of SS 433
during an eclipse, we can extract the spectrum of the very hot central source,
which is covered in eclipses, from the combined light of the system.
The combined light contains also the radiation
from the surrounding gaseous shell and disk. We assume that the extracted spectrum
is an extension of the thermal bremsstrahlung continuum that is radiated by
the jet, its temperature being about 2.3$\cdot10^8$ K (Yuan et al., 1995).
The slope of such a spectrum in the wide wavelength range covering near-UV,
optical, and near-IR domains is similar for temperatures exceeding 10$^6$ K.
The extracted continuum can be used as a calibration standard to measure
the extinction. The problem of extinction can be solved by comparison of the
observed extracted spectral energy distribution to a calculated energy
distribution for a black body with $T_{eff} = 10^6$ K.

\PZfig{10cm}{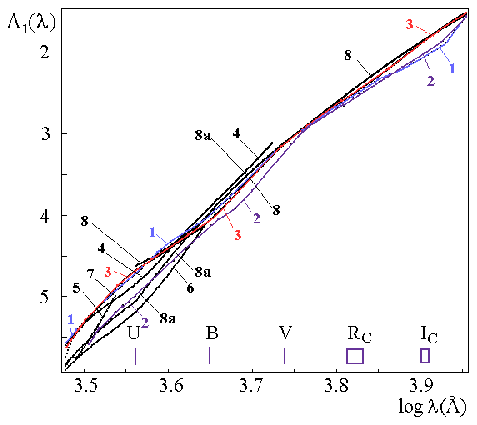}{Curves of the interstellar extinction law
measured in different studies. 1: Sudzius (Straizys, 1977);
2: Schild (1977); 3: Savage \& Mathis (1979); 4: Prevot et al. (1984)
for the SMC; 5: Nandy et al. (1989) for the LMC; 6: Koornneef \& Code (1981)
for the LMC, from observations with the IUE observatory; 7: Seaton (1979); 8: Howarth (1983)
for the Galaxy; 8a: for the LMC. The extinction curve by Savage \& Mathis (1979) (No.~3),
plotted red, was chosen for this study.}

\section{OBSERVATIONS AND DATA REDUCTION}

To analyze spectral energy distributions, I used multicolor CCD $UBVR_CI_C$
photometry of SS 433 in two eclipses centered at 2003 October 1 (JD 2452914.77;
$\psi = 0.97$) and 2007 October 6 (JD 2454380.33; $\psi = 0.99$) from our
collection mentioned in Sec.~2. The observations were made with the SAO 1-m
Zeiss reflector and a photometer equipped with an EEV 42-40 chip and
standard filters.
Observations of the eclipse on 2007 October 6 were a part of a big campaign and
accompanied with extensive spectroscopy using the Russian 6-m BTA
telescope and the Japanese 8-meter Subaru telescope. The results of this campaign
were published by Kubota et al. (2010). During spectroscopic observations with the
Subaru
telescope, a series of direct CCD frames with the $V$ band filter were acquired
in order to point the star into the spectrograph slit. I reduced these frames and
also used them for the analysis. Additionally, I used the BTA and Subaru
spectroscopy to estimate the emission-line contribution in the photometric
bands. The Subaru/FOCAS spectra are from four nights between 2007
October 6 and 10 and cover the wavelength range of 3750 $-$ 5250 \AA\ with a
dispersion of 0.37 \AA\ per pixel. The BTA/SCORPIO spectra were taken on four nights
between 2007 October 4 and 7 and cover the range of 3944 $-$ 5705 \AA\ with a
dispersion of 0.86 \AA\ per pixel.

\PZfig{10cm}{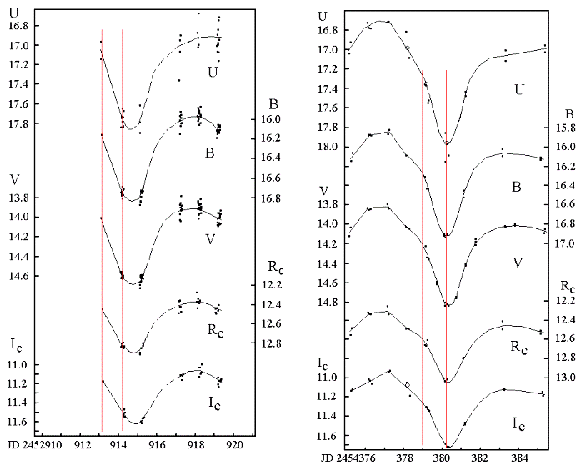}{The light curves of two eclipses, on 2003 October 1
(left) and on 2007 October 6 (right) in the $UBVR_CI_C$ filters (arranged in the
order of filter wavelength from top to bottom). The vertical red lines mark
the phase interval of the ingress of the hot central light source.
The light loss between these phases is used to extract the spectral energy
distribution of the hot source with the difference method.}

The CCD photometry was reduced in a standard way using bias and flat field frames.
Dark frames were not applied because of deep freezing of the CCD chip with
liquid nitrogen to a temperature of $-$133\deg C. The thermal noise is negligible
at this temperature. The nearby star Lyuty 9 (GSC 0071-00141), the southern star of the rhomb, was
the comparison star. The star has a faint red companion at 6\farcs0 NE; its
contribution to the combined light is significant in the $V, R$ and $I$ bands
in the cases of poor seeing. On such occasions, the comparison-star pixels
affected with the companion were eliminated from the star profile,
and their intensities
were replaced with average-profile intensities when computing the integral light.
Extraction of images was performed with my software WinFITS in the aperture
mode, with a star-profile correction.
The $UBVR_CI_C$ magnitudes of the comparison star are the following: 15.073,
14.581, 13.431, 12.973, 12.383. Additionally, four check stars
were used. The uncertainty of the measurements is about 0\fmm02 $-$ 0\fmm04
in the $BVR_CI_C$ filters,
and about 0\fmm1 in $U$ filter. Generally, several observations were done on a
night. The light curves of the two above-mentioned eclipses are shown in Fig.~7
for all the filters. In a few cases when nightly differences of measured
magnitudes were sufficiently large, I co-added frames with matching star
images and remeasured the nightly-sum frame. Such observations are marked with
circles in the Figure.

As follows from the X-ray observations (Fig.~5), the ingress of the jet base begins at
$\phi \approx$ 0.905, the egress ends at $\phi \approx$ 0.095 and the eclipse of
the jet is partial. Near the jet base ingress, the optical light decline is
most rapid.
To extract the spectral energy distribution of the hot light source using
the difference method, we should include phases $\phi$ = 0.905 of both
eclipses in the phase interval where the light loss of the hot source
happens. These intervals are shown in Fig.~7 with vertical lines. In the eclipse
on 2007 October 6, both ingress and egress were observed. The shape of the
eclipse was symmetric, the light curve is well compatible with its
reflected one. Therefore, I used both egress and ingress to measure the
light lost in the eclipse. I interpolated magnitudes to include phases of
the jet base contact. To compare the energy distribution of the light lost in
the eclipse to the Planck energy distribution, we should transform
magnitudes into physical units of $erg\cdot cm^{-2}\cdot s^{-1}\cdot A^{-1}$.

\section{ABSOLUTE CALIBRATION OF THE $UBVR_CI_C$ SYSTEM}

It is known that the zero point of the photometric $UBVR_CI_C$ system is related to
A0V stars. It is accepted by definition that all unreddened color indices
of such star are zeros.
Spectrophotometry deals with monochromatic
intensities, i.e. intensities are measured in the physical units, and
an element of spectral dispersion (pixel) is so small that intensity
varies insufficiently between nearby elements. In the wide-band photometry,
the width of transmission curve is usually of thousands of Angstr\H{o}ms, and stellar
magnitudes are heterochromatic. In the spectral energy distribution measured
with wide-band photometry, each magnitude corresponds to the mean wavelength
of the device transmission curve $r(\lambda)$, calculated as:\\

$\lambda_0 = \frac{\int r(\lambda) \lambda d\lambda}{\int r(\lambda) d\lambda}$
\ \ \ \ \ \ \ (8)\\

\noindent
(Straizys, 1977). The heterochromatic extinction is equal to the monochromatic
extinction at the effective wavelength calculated as:\\

$\lambda_{eff} = \frac{\int I(\lambda) r(\lambda) \lambda d\lambda}
{\int I(\lambda) r(\lambda) d\lambda}$ \ \ \ \ \ \ \ \ (9)\\

\noindent
(King, 1952). $I(\lambda)$ is the monochromatic spectral energy distribution.
Equation (9) means that the intensities measured with wide-band filters correspond,
in the energy distribution, to $\lambda_{eff}$ but not to $\lambda_0$.
Replacing $\lambda_0$ with $\lambda_{eff}$ for each filter and calculating
intensities between filters for the wavelength set of the extinction curve
with the linear interpolation method, I tried to reduce the solution of
the photometric problem to a spectrophotometric one.

Straizys (1977) used the calibration of $UBV$ magnitudes from Straizys
\& Kuriliene (1975), based on the Vilnius photoelectric system. These calibration flux
densities for a zero-magnitude A0V star are given in the fourth column of
Table 2. The mean wavelengths $\lambda_0$
for the filters given in column 2 of this Table were
calculated by me via eq. (8) using filter transmission curves
presented in the digital form by Moro \& Munari (2000). Moro \& Munari
give Vilnius coefficients along with another version of coefficients, which
became popular and are called the photoelectric USA version (Matthews \& Sandage,
1963). These $UBV$ data along with the Cousins $RI$ calibration coefficients
are given in Column 5 of Table 2. Note that the reduction coefficients
of the Vilnius and USA systems differ by 5.7 percents in the $U$ band and by
8.6 percents in the $B$ band, just in the spectral region where the strong
multiple Balmer absorptions of an A0V star
 and its Balmer jump are located. Such a
big uncertainty is inadmissible for SS 433 calibration.

Therefore, I examined the reduction coefficients of the $UBVR_CI_C$ system using
Moro \& Munari transmission curves with the modern spectral energy distribution
of the A0V star Vega published by Bohlin \& Gilliland (2004).
The $V$ magnitude of 0.026 $\pm$ 0.008 is established for Vega in the cited paper, and
the absolute flux level is 3.46$\cdot$10$^{-9}$ erg cm$^{-2}$s$^{-1}$ at
$\lambda$ 5556 \AA. The reduction coefficients were calculated as
average values of the Vega flux density weghted with the filter transmission curve, as
follows from the formula:\\

$f_\lambda = k \cdot \frac{\int FD_V(\lambda) r(\lambda) d\lambda}
{\int r(\lambda) d\lambda}$, \ \ \ \ \ \ \ (10)\\

\noindent
where $FD_{V}(\lambda)$ is the spectral energy distribution of Vega
(flux density function), and $k = 10^{0.4\cdot0.026} = 1.024$ is the correction
for the difference between the Vega $V$ magnitude and zero.
The calculated reduction coefficients
are presented in the last, sixth column of Table 2. These coefficients may be used for
normal stars with atomic absorption spectra. But they should not be used for
cool stars with wide absorption bands of molecular emission, and certainly not for
stars with such a big extinction as that of SS 433. Straizys (1977) noted that
calibration coefficients for the wavelength between filters' mean wavelengths
in the infrared range can be calculated by means of interpolation.
However, my tests in the $U$ and $B$ spectral range demonstrate that it is not true
at short wavelengths due to multiple deep Balmer absorptions and the Balmer jump.
Thus, for SS 433, I calculated special reduction coefficients
not for filter transmission curves but for actually measured light, according
to the expression:\\

$f_\lambda' = k \cdot \frac{\int FD_V(\lambda) I_{SS}(\lambda) r(\lambda) d\lambda}
{\int I_{SS}(\lambda) r(\lambda) d\lambda}$, \ \ \ \ \ \ \ (11)\\

\noindent
where $I_{SS}(\lambda)$ is the first approximation for the SS 433 energy distribution
calculated by interpolation, with normal reduction coefficients.

\PZfig{8cm}{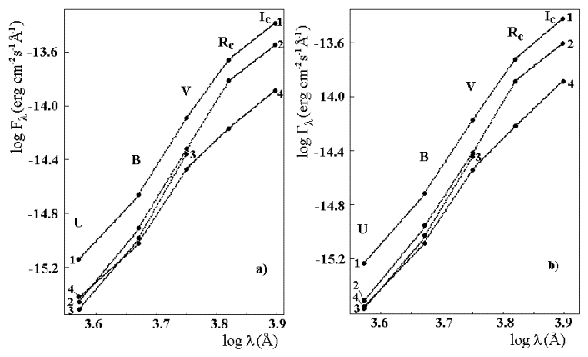}{The spectral energy distributions of SS 433 in
two eclipses, on 2003 October 1 (a) and on 2007 October 6 (b), not corrected
for interstellar extinction. 1: the near-contact distribution, jet base included;
2: the distribution near the eclipse center, jet base covered; 3: the energy
distribution in the $UBV$ bands near the eclipse center, with the contribution from
emission lines eliminated; 4: the energy distribution for light lost in the
eclipse.}

The results are given in Table 3, where the effective wavelengths of measured
light are given in Column 2, their logarithms are in Column 3, and the "A0V-star
zero-magnitude" reduction coefficients for these effective wavelengths are in
Column 4. Comparison of Columns 2 in Tables 2 and 3 shows that the effective
wavelengths in the short-wavelength filters differ from the mean wavelengths by
about 200 \AA, and this effect should be taken into account for SS 433.

\PZfig{8cm}{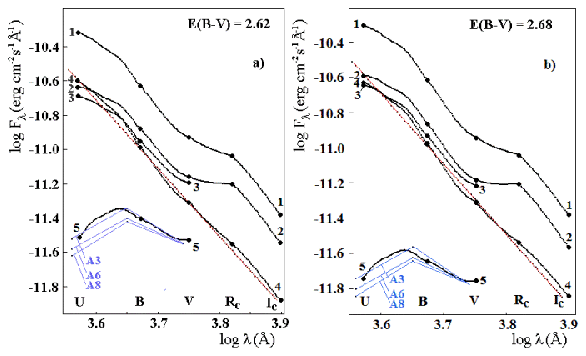}{The spectral energy distributions of SS 433 in
two eclipses, on 2003 October 1 (a) and on 2007 October 6 (b), corrected for
interstellar extinction. Numbers 1 $-$ 4 indicate the same energy distributions
as in Fig. 8. 5 is the extracted energy distribution of the donor.
The straight red line is the Planck energy distribution for a black body with
T$_{e}$ = $10^6$ K. The blue curves are unreddened photometric energy
distributions for the stars HD 12027 (A3 III), HD 240296 (A6 III), and
HD 12161 (A8 III).
}

The spectral energy distributions of SS 433 in the two eclipses, after such a
correction but not yet corrected for interstellar extinction, are
shown in Fig.~8.
The results of energy-distributions correction for interstellar extinction
made individually for both eclipses are shown in Fig.~9. In Fig.~8 and 9,
different energy distributions are plotted, those of light near
the jet-base contact with the jet base included,
those of light near the center of eclipse with
the jet base covered in the eclipse, and those of light lost in the eclipse.
The distribution of the lost light is calculated as the difference of
distributions with the jet base included and excluded.
The correction for interstellar extinction was made by
choosing a color excess value to get the best fitting, by eye, of the lost-light
distribution (No. 3) with the black-body distribution for \hbox{T$_{e}$ = $10^6$ K}
(the straight red line in Fig.~9). In such fitting, the level of black-body
distribution was a free parameter, so a color excess was chosen only to fit its
slope. With the photometry of 2003 October 1 eclipse, the best-fit color excess
is 2\fmm62, and for 2007 October 6 eclipse, it is 2\fmm68.
The average is $E(B - V)$ = 2\fmm65 $\pm$0\fmm03.

It is important to note that the strong excess in the  $R$ and $I$ bands described
above is revealed in every energy distribution of SS 433 being compared to
the distribution of light lost in the eclipse. It is seen both in Fig.~8 in
the reddened data and in Fig.~9 in unreddened data.
The contribution of
this excess can be easily extracted from the combined light. For this purpose, we
extrapolate linearly the short-wavelength part of the combined-light distribution
to the $R$ and $I$ bands and subtract the extrapolated values. However, the excess
veils such a faint light source as an A-type star and makes its extraction
impossible in the $R$ and $I$ bands.

The shape of the lost-light energy distribution, corrected for reddening with the
determined $E(B - V)$, is nearly a straight line.  This fact confirms
my assumption on the nature of this light source as an optical tail of the energy
distribution radiated by the jet bases. The maximum of this energy distribution is
located far in X-rays. The individual points of the lost-light distribution
indicate small deviations from the straight line in different filters (Fig.~9a,b),
probably due to systematic errors in comparison-star magnitudes, uncertainties
of the chosen extinction law, or inconsistency of the general law with the individual
law in the particular sky direction, etc. Thus, the individual extinction law can be
derived as the difference between the observed lost-light energy distribution and
the selected calculated Planck energy distribution divided by the average of
reddening.

If we adopt the light contribution of an A4 $-$ A8 donor between 0.29
and 0.43 in the eclipse in the $V$ filter (Hillwig \& Gies, 2008), and taking
into account the earlier described parameters and uncertainties of distance and
extinction, the absolute magnitude of the donor will be limited to
$-6\fmm28 \le M_V \le -5\fmm40$. In the spectral type $-$ absolute magnitude diagram
(Fig.~10), the star occupies the uncertainty region plotted as a green rectangle.
The radius of such a star can be calculated using the Stephan $-$ Boltzmann law,
$L = 4\pi R^2 \sigma T^4$, or, in solar units,\\

$\frac{L}{L_\odot} = (\frac{R}{R_\odot})^2 \cdot (\frac{T}{T_\odot})^4$.
\ \ \ \ \ \ \ (12)\\

\noindent
With this formula, the radius of A-type donor can be estimated as
50 - 80 R$_\odot$. Naturally, with such a companion, we have no doubt in the
"thin-jet" hypothesis when treating the X-ray eclipse.
Otherwise, it is impossible to invent a collimation mechanism for the jet with
the thickness of dozens of solar radii and with the collimation
angle of 1.0 $-$ 1.4 degrees by a compact object.

\PZfig{11cm}{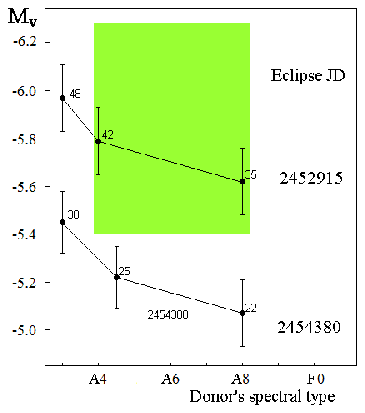}{The spectral type $-$ absolute $V$ magnitude diagram.
The green rectangle is the region of spectroscopic solutions. The points with broken lines
are photometric solutions for the two eclipses. A number
near each point is the contribution of the donor A star in the combined light
of the system in eclipse}.

There is another question of interest. Can we detect, extract, or measure
the contribution of the donor star during an eclipse from multicolor photometry
knowing the possible range of its spectral type, A4 $-$ A8? The spectral energy
distributions of SS 433 during eclipses (curves 2 in Fig.~9a,b) contain,
besides the radiation from the donor, also the radiation
from the hot source (possible jet) eclipsed partially and the emission from
the surrounding gas envelope, which may include also an equatorially expanding
extended disk around the system described by Barnes et al. (2006, Fig.~1).
Barnes et al. show that the donor is visible
during the eclipse, near the precession phase $\psi \approx 0$, being not shielded by the
extended disk. Using the Subaru and BTA spectra, I measured the emission
contribution in the $U,B$ and $V$ filters in the eclipse of 2007 October 6; the
results are given in Table~4. The averages of these contributions to the $UBV$
filters were subtracted from the intensities of light during eclipses (curves 2,
Fig.~8 and 9), to get a pure continuum distribution in eclipses (curves 3).
Note that the same emission-line contributions were subtracted also from the energy
distribution of the 2003 October 1 eclipse because those observations were not
accompanied by spectroscopy.

The unreddened $UBV$ colors of HD 12027
(A3 III), HD 240296 (A6 III), and HD 12161 (A8 III) were taken from Table 2 in
Jacoby, Hunter \& Christian (1984) as examples of A-star colors. I plotted the spectral
energy distributions of these stars in Fig.~9 as blue curves, for comparison,
so that they intersect in the $V$-band range; the $V$-band intensity
was used as a free parameter. Then I removed some portions of hot-source energy
distributions (curves 4) from the eclipse-center energy distributions corrected
for emission (curves 3), to find, in the residuals, an energy distribution
resembling that of an A star (see curves 5 in Fig.~9).
This is easy to do because a real star with a large Balmer jump is present
in each of the two energy distributions.
Accurately adjusting the subtracted hot-source energy distribution (curve 4) and
the $V$-band intensity of the template, one can fit the residual distribution with any star,
from A3 to A6. In the case of best fitting, $V$-band intensity of the sample
corresponds to unreddened visible $V$ magnitude of the donor. Certainly, the
visible magnitude, absolute magnitude, and donor's contribution
to the combined light (which is calculated for the $V$ band)
depend on the energy distribution of the template chosen for fitting and on the
amount of hot-source radiation. The results of residuals are shown
in Fig.~10 for each eclipse as points with error bars; the contribution
of the A-type donor to the $V$ band, estimated from each fitting and expressed
in percents, is indicated by numbers near each point. It appears from Fig.~10
that the absolute magnitudes derived from photometry are systematically fainter
than those derived from spectroscopy. Additionally, the derived absolute
magnitudes of the donor change from eclipse to eclipse, so that their
error bars do not overlap. The nature of these systematical differences
is not yet clear. Using the two eclipses, we find the following absolute-magnitude
limits: $-5\fmm9 \le M_V \le -5\fmm0$, for the donor spectra in the A4 $-$ A8
range.

\section{THE MASS $-$  LUMINOSITY RELATION}

The Russell-Vogt theorem states that if we know a star's mass and
its chemical composition, then, using laws of physics, we can determine all
of its other properties, such as luminosity, radius, temperature and density profiles,
and find how these properties change with time (Massey \& Meyer, 2001).
A. Eddington was the first to demonstrate that radiative diffusion in stars required
a dependence of the stellar luminosity on mass, $L \sim M^4$. Since Eddington
times, stellar luminosities are known from evolutionary modeling. Modeling
shows a good mass $-$ luminosity relation for main-sequence stars with hydrogen
burning in their centers. But low-mass stars, like those in globular clusters,
can reach high luminosities during late stages of their evolution, for example,
the stage of a red giant with a degenerate helium core or the stage of an AGB star
with a degenarate carbon core. Such stars deviate far from the general
relation though for a comparatively short time. Other unaccounted effects
in this theory, like rotation or binarity, can lead to violations of the relationship.

Stellar masses can be determined from observations if we apply dynamic methods
to visual, eclipsing, or spectroscopic binaries, mostly detached systems.
Certainly, the knowledge of accurate distances and luminosities is needed
for such systems. Using this method for semidetached or contact binaries
faces the problems described in this paper for the case of SS 433.
Besides, it is possible to estimate a star's
mass from high-resolution spectroscopy and stellar-atmosphere modeling.
Fitting the observed line profiles to model ones gives the effective
temperature $T_{eff}$ and surface gravity $g$. If the star's distance and
luminosity are known, one can calculate its radius from the Stephan $-$ Boltzmann law
(cf. eq.(12)), and then its mass from the relation $g \sim M/R^2$
(Massey \& Meyer, 2001).

\PZfig{8.6cm}{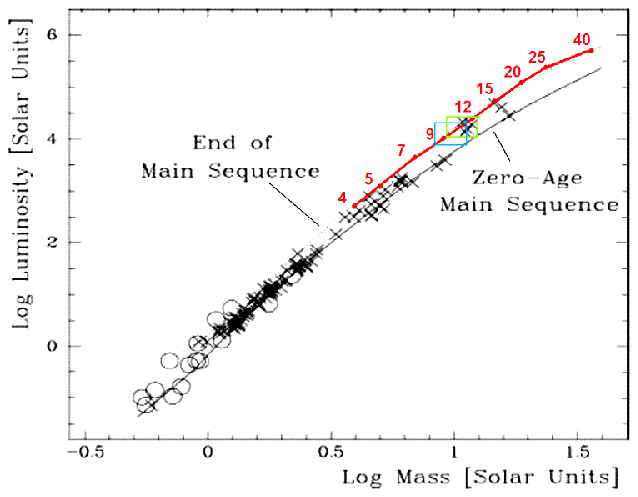}{A fragment of the mass $-$ luminosity
diagram by Massey \& Meyer (2001, Fig.~1) constructed for components of
binary stars. The theoretical mass $-$ luminosity relation
derived by the author for A stars using the models by Schaller et al. (1992)
is also plotted (red points and lines).
There is an agreement with observations of binary components.
The localization of SS 433 A-type donor is marked with two error boxes,
the green box being the localization revealed from spectroscopy and the blue box,
from photometry.}

I analyzed the mass $-$ luminosity relation for A-type stars using
the evolutionary
models by Schaller et al. (1992), specifically their evolutionary tracks for
a nearly-solar chemical composition (Y = 0.30, Z = 0.020, their Fig.~1).
Massive and luminous stars became A stars on their tracks to red giants.
The evolution of massive stars to the red side of the C-M diagram
goes at a constant
luminosity, and hence the  theoretical mass $-$ luminosity relation is determined
correctly, with an insignificant dispersion. The theoretical mass $-$ luminosity
relation for A stars is presented in Table 5. In Fig.~11, this relation is
plotted over a fragment of the mass $-$ luminosity diagram presented in Fig.~1
of Massey \& Meyer (2001), which was constructed for components of binary stars.
The relation for A
stars is plotted as red points and a red broken line; the red numbers are initial masses of
the models. Along the abscissa, current masses of
A-star models are plotted: stellar models were
computed with the mass loss taken into account. In the luminosity ranges of interest,
the mass loss of single stars that become A stars varies from 0.013 $M_{\odot}$
for the initial mass of 9 $M_\odot$ to 0.10 $M_{\odot}$ for 12 $M_\odot$.
Figure 11 shows that the A-star relation coincides with the location of stars
at the end of the main sequence, whereas the stars on the zero-age main
sequence have larger masses for their luminosities. Our comparison of the theory
with observations of star-system components reveals a good agreement.

SS 433 is not a single
star, and the donor is losing mass at a higher rate than
a single star, but the current bolometric luminosity of the donor should reflect
its curent mass.
Principally, the bolometric absolute magnitude does not depend
on the star's spectrum (or on  $T_{eff}$ on its surface), it depends on the
energy release in the stellar interior. Eventually, howerer, the energy release depends
on stellar mass.

When using the theoretical relationship for A stars to estimate the donor mass in
SS 433, the question arises if this star has an internal structure similar
to normal single stars with the same spectral types and luminosities. Actually, the
volume of a star in a binary system is limited by the critical equipotential
surface. A component of a binary evolving to the stage of degenerate helium
core would not be able to become a red giant because its expanding envelope
will overflow to the secondary component or flow away from the system.
Such phenomena are observed in SS 433. A star with a forming helium
core may have excess luminosity compared to a single star of the same
mass. Using the relation for single stars, we will overestimate the
mass of such a component.
A common-envelope phase in a binary with an expanding component is also possible,
and this can be just the case for the system of SS 433.

For masses exceeding 7 $M_\odot$, loops of evolutionary tracks
occuring in the He-burning phase get into the region of A stars at somewhat higher
luminosities than the luminosity on the way to red giants. The He-burning phase
on such a loop seems improbable for the SS 433 donor. However, using
the mass $-$ luminosity relation for A stars in such a case, we will also overestimate the mass
of the SS 433 donor. This means that if we use the mass-luminosity relation for A stars
derived using tracks to red giants but the real A star is at a later evolutionary
stage and its envelope is limited by the binary's equipotential surfice,
then its mass will actually be lower, and lower will be the mass of the compact object.

The donor's mass determined from this relation, in agreement with its
light contribution estimated spectroscopically, falls in the range from 9.4 to 12.5
$M_\odot$, while the mass estimated photometrically is between
8.3 and 11.0 $M_\odot$. The error boxes of the parameters derived for the donor
are shown in Fig.~11 as rectangles. The green rectangle corresponds to
the spectroscopically determined contribution in eclipse, and the blue rectangle
corresponds to that determined photometrically. The location of the error boxes
is in agreement with models for evolved stars that leaved the main sequence.

The mass $M_X$ of the compact object can be calculated
using the formula $M_X = q M_A$, where $M_A$ is the mass of the A-type donor and
$q = M_X/M_A$ = 0.15 is the mass ratio known from observations of
the jet eclipse in X-rays.

\PZfig{11cm}{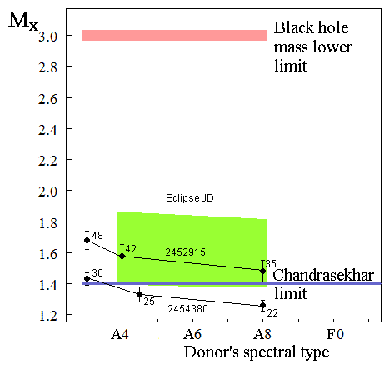}{Mass estimates of the compact object based
on the estimates of A-type donor luminosities. As in Fig.~10, the green
rectangle is the region of spectroscopic solutions. Points with error bars
connected with broken lines are photometric solutions for the mass
revealed from the photometry of the two eclipses. A number printed
near each point is the $V$-band contribution (in percents) of the A-type donor to the
the combined light of the system in eclipse.}

Now, having in mind all these considerations and using the small bolometric corrections,
in the range between $-$0\fmm05 and +0\fmm09, respectively for A4 I$-$III and
A8 I$-$III, we translate all the regions and points shown in Fig.~10
(the spectral type $-$ $M_V$ diagram) to the donor's spectral type $-$ compact-object mass diagram
for SS 433 (Fig.~12).

Fig.~12 demonstrates that all mass estimates are concentrated near the
level of the Chandrasekhar limit established for zero-age neutron stars,
1.4 $M_\odot$ (blue line), but essentially below the lower mass
limit known for stellar-mass black holes, \hbox{3 $M_\odot$}
(red line).
Note that wrong identification of the donor's evolutionary stage
would lead to smaller mass estimates for the compact component,
and its mass will still be below the Chandrasekhar limit.

Our final mass estimates for the compact object in SS 433 are 1.64 $\pm$0.23
$M_\odot$ for the donor's light contribution in the eclipse near T$_3$
derived spectroscopically and 1.45 $\pm$0.20 $M_\odot$ for
that derived photometrically and based on the donor's spectral type estimated
from spectroscopy as A4 $-$ A8 I$-$III.

To check if the masses of the components of SS 433 agree with
laws of physics, let us calculate the amplitudes of radial-velocity curves $K_A$ for the
A-type donor and $K_X$ for the compact object from the formulae
of Keplerian dynamics:\\

$f_X(M) = \frac{M_X^3 \cdot sin^3i}{(M_X + M_A)^2} =
10385 \cdot 10^{-11}(1 - e^2)^\frac{3}{2} \cdot K_A^3 \cdot P$; \ \ \ \ \ \ (13)\\

$\frac{K_A}{K_X} = \frac{M_X}{M_A}$;  \ \ \ \ \ \ \ \ (14)\\

\noindent where $M_A$ and $M_X$ are respectively the masses of the A-type donor and the compact object;
$e$ is the orbital excentricity, assumed to be zero; $i$ is the
orbit inclination, 78\deg.81; $P$ is the orbital period, 13$^d$.08223.
The results are the following: $K_A =$ 26 km s$^{-1}$,
$K_X =$ 170 km s$^{-1}$. Thus, the velocity-curve amplitude of the A-type
donor is really 1.5 times smaller than the same parameter in Kubota et al.
(2010), even that corrected for the heating effect, 40 km s$^{-1}$.
At the same time, the velocity-curve amplitude of the compact component
is well in compliance with the amplitudes revealed from
the He II emission line, 175 km s$^{-1}$ (Fabrika \& Bychkova, 1990), or from the CII
emission blend, 162 km s$^{-1}$ (Gies et al., 2002b).

Let us now  conversely calculate the magnitude of the donor star using the mass
of 4.3 $M_\odot$ for the compact object from Kubota et al. (2010) and
3.0 $M_\odot$, the mass of a hypothetical black hole at the lower limit of
black-hole masses. With the mass ratio $q$ = 0.15 and the mass $-$ luminosity
relation deduced from Shaller et al. (1992), I estimated
masses, absolute $V$ magnitudes, and $V$-band visible reddened
magnitudes of the A-type donor stars as respectively
28.7 and 20 $M_\odot$, $-$9\fmm13 and $-$8\fmm25, 12\fmm67 and 13\fmm55.
Note that SS 433 has a maximum normal brightness 13\fmm9 $-$
14\fmm0 in the $V$ band, and its brightness reaches 13\fmm55 only in outbursts.
This is the combined brightness of all components and light sources in the binary.
Thus, black-hole masses are in a strong conflict with the results of photometry.

\section{DISCUSSION}

The general parameters used in this paper, such as distance, extinction, A-star
contribution during eclipses near $T_3$ precession phase are no surprise.
They were repeatedly examined in different studies. The contribution of the A-type donor
measured in this paper, 32 $\pm$ 10 percents
in the $V$ band, photometrically corresponds to a
magnitude between 15.7 and 16.4. This range is in conflict with
the magnitude of 17.35 in the $V$ band measured in an eclipse, at the $T_2$
phase, by Henson et al. (1982). However, this contradiction can be easy understood if
we remember the circumbinary equatorial expanding and precessing disk,
well illustrated by Barnes et al.
(2006), which covers the system partly in the precession phases near $T_1$ and
$T_2$. The known dependence of eclipse depth on the precession phase with
the amplitude of 0\fmm40 can be explained with this disk, too. The eclipse
observed by Henson et al. was monitored also by Gladyshev et al. (1987), and
it is known that, 0$^d$.5 before and 0$^d$.5 after the observation by Henson et
al., the brightness of SS 433 was extraordinary high
for those orbital and precession phases, so the object was in a
high state. Probably, a time was caught when the circumbinary disk was
extraordinary thick and covered the donor totally. This event could
coincide with a short-duration weakening of the jets. Such an event
was unique but observationally well established.

The nature of the circumbinary disk  remains unclear yet. It is observed in radio
bands as a wind of rapidly varying shape at the distances of about
200 $-$ 250 AU and is resolved (Paragi et al., 1999). This structure is
perpendicular to radio jets. This radio structure is discussed in detail
in the review by Fabrika (2004); see additional references there.
The disk may be formed by gas flowing out from the external Lagrangian point L2.
It is unclear why this gas, which covers the central jet bases, does not form an
absorption-line spectrum in the optical bands.

With the A-star of 32 $\pm$10 percents in the $V$ band
in minimum light, its contribution is only 10 - 20 percents in maximum
light. It is clear from an inspection of Fig.~2, where the light curve near $T_3$ has a
$\beta$ Lyr-like shape with a large-amplitude "ellipsoidal effect"
and a secondary eclipse 0\fmm4 deep, that neither the
ellipsoidal effect no the secondary eclipse can be related by the donor because
of its small light contribution. The main bright source in the $V$ band
is the very hot one, covered partially in the primary eclipse. Then its
contribution in maximum light is up to 77 $-$ 87 percents, taking into
account the 3-percent contribution of the emission-line-radiating envelope.
If this is a hot disk or another rapidly rotating body, it also cannot form either
the ellipsoidal effect or the deep secondary eclipse. Otherwise,
the source may be a star filling the Roche lobe of the compact component and
having an elongated shape due to gravitational influence of the donor. Thus, the
hypothesis of a "supercritical accretion disk" in SS 433 is a myth, and it is
not confirmed with the photometry. It is not correct to fit the optical light
curves by a model of a thick precessing disk plus a gravitationally
distorted OB star, as that was done by Antokhina \& Cherepashchuk (1985, 1987).
Moreover, we do not see any details in the photometry that could
be treated as external contacts of a
bright star filling the Roche lobe of the compact component, though know
that the eclipse of the Roche lobe is total due to the small ratio, $q = 0.15$.
No contribution of such a star, which may have an energy distribution
different from that of the very hot source, is also detectable in
multicolor photometry. It is small compared to the A-type donor distribution.

In such a case, we may expect that the light distribution of the hot source in
SS 433 is formed only by jet bases and holes or nozzles the jets come
from. The "ellipsoidal effect", the precession light variations, and the
"secondary eclipse" of the very hot light source are present only due to
effects of visibility of the nozzle depths and jets' bases.
Near precession phases $T_1$ and $T_2$, the dispersion of light
curves increases strongly (Fig.~2); the causes of this phenomenon
may be both the system being covered by the
nonuniform structure of the hypothetical circumbinary disk and
variations of visibility conditions of nozzle depths due to uneven and rough
structure of nozzles' edges. The hypothesis that the light curves of
SS 433 could be explained by the motion of two bright hot spots at the points
of emergence of the jets (or jet nozzles), while the contribution of the disk
or its envelope to the total brightness was small, was first put toward by
Lipunov \& Shakura (1982).

Another important factor of forming the optical light can  be
variable brightness of jets. Hydrogen burning on the surface of a neutron star
is very unstable, because a critical mass should be accumulated repeatedly
for the ignition. Repeated ignitions may be the cause of forming the bullets.
Appearance of bullets in the moving Balmer lines means that the matter of
jets is thrown out in portions. The connection of the bullet formation to the
optical variations has not yet been studied. However, this effect can cause deviations of
individual minima from the linear ephemeris found in my study.
Additionally, the jets are driven by tidal waves responsible for nodding motion.
Fourier analysis of our photometry also shows a significant wave with an
amplitude of about 0\fmm2 and the same period as that of nodding motion
of the moving $H_\alpha$ components in the spectra, 6$^d$.2887 (Goranskij et
al., 1998a).

I assume that, in the case of a contact system in SS 433 having a neutron star in
the center of a component filling the Roche lobe, the products of hydrogen
burning on the surface of the neutron star are channelling out of the star
through the nozzles throwing most of the thermonuclear energy into
space. Therefore, the surface of such a star with the neutron star inside
can be heated insufficiently to contribute to the common light of the
system. I think that cases of supercritical accretion on a black hole
and of hydrogen burning on the surface of a neutron star differ in
the chemical composition of matter erupted in jets. Indeed, no
thermonuclear burning is possible if the matter falls into a black hole.
I should be also reminded that Lipunov \& Shakura (1982) demonstrated the
liberation of energy through the accretion process on a neutron
star to be the only process capable of explaining the energy release of SS 433.

The conclusions of this study are strongly dependent on the mass ratio $q$ = 0.15
established from X-ray observations of the Ginga satellite. I suppose that
it is a very reliable result because the eclipse contacts were observed not
only in continuum radiation but also in the moving emission blend of Fe XXV/Fe XXVI
radiated by the jets. This circumstance permits to
correctly localize the eclipsed source. The contacts and width of the X-ray eclipse
are confirmed with the X-ray observations in continuum. Another confirmation
of the mass ratio being so small is the radial velocity of the stationary H$_\alpha$ line,
which is displaced by 90 degrees in orbital phase. The maximum recessive
velocity in H$_\alpha$ is observed in the inferior conjunction of the donor,
i.e. in eclipse. The equivalent width of H$_\alpha$ line increases in eclipse
due to deep eclipse in continuum, but the intensity of the line does not show
any eclipse. This means that the line is radiated in an extensive and expanding
envelope.

Goranskij et al. (1997) find a natural explanation for this phenomenon. In this
system, a part of the envelope is located in a shadow of the companion.
The companion covers the radiation of the hot source for the shadow cone,
gas in the shadow rapidly recombines, and it cannot be excited by the hot
source for the eclipse duration. The opposite side of A-type donor,
which has strong and wide Balmer and Lyman absorption lines, is not able to effectively
excite the hydrogen in the shadow. The angle of the shadow cone, the
volume of neutral gas in the envelope, the lost emission intensity, due to
the shadow, in the line profiles, and the amplitude of the velocity curve based on
mean-profile intensity variations depend strongly on the distance
of the hot exciting source from the donor's surface, and therefore
on the mass ratio $q$. My simple profile modeling with the Monte-Carlo method
shows that the observed amplitude of 160 km s$^{-1}$ is best fitted
with the mass ratio $q$ = 0.15. Thus, the displacement of the H$_\alpha$ radial
velocity curve is an additional argument in favor of low mass ratio and low
mass of the compact companion.

Kubota et al. (2010) try to disprove this argument ristricting it only to "wind
evacuation", i.e. to covering the wind cone by the donor, with the wind being
formed by the accretion disk. They thought this covering led to an
anisotropic wind. Their arguments against the shadow hypothesis are weak.
The first one is that the radial velocity amplitudes of H and He I have to
exceed those of the accretion disk (traced by He II line) since the disk powers
the wind. The presence of such a disk has not yet been proven, the most questionable
is the hypothesis of forming the He II line by the disk. The behavior of the He II line
in eclipse described above does not confirm its relation to the disk. No partial
or total eclipses in H and He I emission have been observed yet. Different
amplitudes of H and He I lines can be explained with different excitation conditions:
the A-type donor has much weaker lines of He I than those of H,
and He I atoms can be excited with a larger probability. The second argument
is related to the accretion gas stream.
If a gas stream drawn in the sketch in Fig.~13 in Kubota et al. (2010)
really exists, it should be
located near the L1 Lagrangian point between the stars and thus should be totally
eclipsed in the phase range wider than that of the jet-base eclipse
because the L1 Lagrangian point is closer to the donor than the jets bases.
Even if located in the drawn place, the gas stream should be eclipsed totally
but in the different range of orbital phases. Nothing of this behavior
was observed. Additionally, $H_\alpha$ has a wide profile with
\hbox{$FWHM \approx$ 1500 \AA\,} corrected for instrumental resolution in the spectra,
so one should find a hypothetical stream component in the profile formed by the
expanding wind with the velocity of about 750 km s$^{-1}$ and exhibiting
wide eclipse.
Recently, Bowler (2009) detected a transient component in the profile of
the $H_\alpha$ emission, subject to Doppler effect with the orbital period,
at a speed of approximately 175 km s$^{-1}$, but this component is not
eclipsed in any orbital phase (the mass he estimated for the compact object is
less than 37 M$_\odot$). Bowler attributes this component to the
accretion disk around a compact companion.

Thus, there exists additional evidence for a very small mass ratio for the components of
SS 433 besides the Ginga X-ray observations in eclipses. Since Ginga
times, new perfect X-ray observatories and devices have been launched into
space and work successfully. One of such observatories is Chandra.
Its HETGS spectrograph has a higher resolution and sensitivity than that of
Ginga. In variance with Ginga, this device resolves the Fe XXV/Fe XXVI blend at
6 keV (Marshall et al., 2002). The Chandra team did not repeat the Ginga experiment
aimed at detecting jet contact times during eclipses, but public releases at the Chandra
Internet site are full with declarations that SS 433 is a black hole binary.
Why not to check Ginga data with the modern Chandra observations?

Another problem which may have an improved solution is the mass $-$ luminosity
relation for A-type stars with high accretion rate caused by the evolution
with the Roche-lobe-limited envelopes. In this study, it was naturally
supposed that A-type stars that had lost their mass from the
envelope evolved as single stars but with a smaller, residual mass. But
it is of interest to solve this problem by modeling. It seems, that this problem is
not so difficult.

As a critical reader of my paper may note, I used, in Section 2, the filter
transmission curves collected by Moro \& Munari (2000) for the original
Johnson and Cousins photometric systems, but not those for the instrumental
$UBVR_CI_C$ bands used in our measurements with the SAO 1-m telescope. Accurate
studies of such curves for the instrumental systems need special devices, like
a monochromator, and techniques of high-precision photometry described by
Mironov (2008).
Additionally, my experience shows that high-precision CCD photometry also needs
special measures to prevent flashes originating from skew light reflection
on the surfaces parallel to the optical axis of the telescope, such as the
Cassegrain blend or cylindrical photometer details, even blackened. For
precise CCD
photometry, it would be preferable to have a special photometric telescope with
a CCD detector in the prime
focus, without any blend. However, we have to use multi-purpose telescope,
not prepared specially for precise photometry. This certainly
affects adversely the accuracy of flat-field calibration and standard
measurements. The absence of spectroscopy in the red
spectral range is another perceptible drawback of this study. We used
the available data as is. Performing absolutely correct photometry of SS 433,
supported with spectroscopy with big telescopes, seems a difficult problem.

The best solution of this problem seems to be spectrophotometry with
a wide and long slit to prevent light losses at the slit, but at the cost
of spectral resolution. Such observations require spectrophotometric standards
in the close vicinity of SS 433. However, establishing such standards is not
a difficult problem.

\section{CONCLUSIONS}

In this paper, it is reliably established on the base of current knowledge of
distance, extinction, mass ratio and contribution of A-type star for SS 433, along
with modern data on stellar photomery, its physical calibration, stellar
bolometric corrections, temperature calibrations, mass $-$ luminosity ratio that
the compact object in the system of SS 433 is a neutron star with a mass
close to the Chandrasekhar limit, 1.4 M$_\odot$. This conclusion follows from
the discovery of an A4 $-$ A8 I$-$III star in the spectrum.
With all possible estimates of A-star's contribution to the combined light
of the system during the eclipse, the mass of the neutron star is in the range
between 1.25 and 1.87 M$_\odot$.
The mass of the neutron star is equal to 1.45 $\pm$0.20 M$_\odot$
solely from multicolor photometry.

To adopt the assumption of a black hole, the basic parameters of SS 433
and main astrophysical data should be radically revised.\\

\textit{Acknowledgments:}

Author thanks A.V. Mironov (Sternberg Astronomical Institute, Moscow University)
for his help in some
questions of stellar photometry and its calibration.

\references

Antokhina, E.A. \& Cherepashchuk, A.M., 1985, {\it Soviet Astron. Letters},
{\it 11}, 4

Antokhina, E.A. \& Cherepashchuk, A.M., 1987, {\it Soviet Astron.}, {\bf 31}, 295

Aslanov, A.A., Cherepashchuk, A.M., Goranskij, V.P., et al., 1993,
{\it Astron. \& Astrophys.}, {\bf 270}, 200

Barnes, A.D., Casares, J., Charles, P.A., et al., 2006, {\it Monthly Not.
Roy. Astron. Soc.}, {\bf 365}, 296

Blundell, K.M. \& Bowler, M.G., 2004, {\it Astrophys. J.}, {\bf 616}, L159

Bohlin, R.C. \& Gilliland, R.L., 2004, {\it Astron. J.}, {\bf 127}, 3508

Borisov, N.V. \& Fabrika, S.N., 1987, {\it Soviet Astron. Letters}, {\bf 13}, 200

Bowler, M.G., 2009, {\it arXiv}:0912.2428

Brinkmann, W., Kawai, N., Matsuoka, M., 1989, {\it Astron. \& Astrophys.},
{\bf 218}, L13

Brinkmann, W., Kotani, T., Kawai, N., 2005, {\it Astron. \& Astrophys.}, {\bf 431},
575

Cherepashchuk, A.M., 1981, {\it Monthly Not. Roy. Astron. Soc.}, {\bf 194}, 761

Cherepashchuk, A.M., Aslanov, A.A., Kornilov, V.G., 1982,
{\it Soviet Astron.}, {\bf 26}, 697

Cherepashchuk, A.M., Sunyaev, R.A., Fabrika, S.N., et al., 2005,
{\it Astron. \& Astrophys.}, {\bf 437}, 561

Crampton, D. \& Hutchings, J.B., 1981, {\it Astrophys. J.}, {\bf 251}, 604

Davydov, V.V., Esipov, V.F., Cherepashchuk, A.M., 2008, {\it Astron. Reports},
{\bf 52}, 487

Fabrika, S.N., 2004, {\it Astrophys. \& Space Phys. Rev.}, {\bf 12}, 1

Fabrika, S.N. \& Bychkova, L.V., 1990, {\it Astron. \& Astrophys.}, {\bf 240}, L5

Fejes, I.,  1986, {\it Astron. \& Astrophys.}, {\bf 168}, 69

Gies, D.R., Huang, W., McSwain, M.V., 2002a, {\it Astrophys. J.}, {\bf 578}, L67

Gies, D.R., McSwain, M.V., Riddle, R.L., et al., 2002b, {\it Astrophys. J.},
{\bf 566}, 1069

Gladyshev, S.A., Goranskij, V.P., Kurochkin, N.E., Cherepashchuk, A.M., 1980,
{\it Astron. Tsirkulyar}, No. 1145

Gladyshev, S.A., 1981, {\it Sov. Astron. Letters}, {\bf 7}, 330

Gladyshev, S.A., Goranskij, V.P. Cherepashchuk, A.M., 1987,
{\it Soviet Astron.}, {\bf 31}, 541

Goranskii, V.P., Fabrika, S.N., Rakhimov, V.Yu., et al., 1997, {\it Astron.
Reports}, {\bf 41}, 656

Goranskii, V.P., Esipov, V.F., Cherepashchuk, A.M., 1998a, {\it Astron.
Reports}, {\bf 42}, 209

Goranskii, V.P., Esipov, V.F., Cherepashchuk, A.M. 1998b, {\it Astron. Reports},
{\bf 42}, 336

Goranskij, V.P., 1998c, {\it Proc. 20-th Conference on Variable Star Research}.
Ed. J.~Dushek, M.~Zejda, Brno, Czech Rep., p.103

Grandi, S.A. \& Stone, R.P.S., 1982, {\it Publ. Astron. Soc. Pacific}, {\bf 94}, 80

Henson, G., Kemp, J., Krauss, D., 1982, {\it IAU Circ.}, No. 3750

Hillwig, T.C., Gies, D.R., Huang, W., McSwain, M.V., et al., 2004, {\it Astrophys.
J.}, {\bf 615}, 422

Hillwig, T.C. \& Gies, D.R., 2008, {\it Astrophys. J.}, {\bf 676}, L37

Hjellming, R.M. \& Johnston, K.J., 1982, In: {\it Extragalactic radio sources.}
Proceedings of the IAU Symposium No. 97, Dordrecht, D. Reidel Publishing Co.
p. 197

Howarth, J.D., 1983, {\it Monthly Not. Roy. Astron. Soc.}, {\bf 203}, 301

Jacoby G.H., Hunter, D.A. \& Christian, C.A., 1984, {\it Astrophys. J. Suppl.},
{\bf 56}, 257

Kawai, N., Matsuoka, M., Pan, H.-C., \& Stewart, G.C., 1989, {\it Publ. Astron.
Soc. Japan}, {\bf 41}, 461

Kemp, J.C., Henson, G.D., Kraus, D.J., et al., 1986, {\it Astrophys. J.}, {\bf 305},
805

King, I., 1952, {\it Astron. J.}, {\bf 57}, 253

Koornneef, J. \& Code, A.D., 1981, {\it Astrophys. J.}, {\bf 247}, 860

Kotani, T., Kawai, N., Aoki, T., et al., 1994, {\it Publ. Astron. Soc. Japan},
{\bf 46}, L147

Kubota, K., Ueda, Y., Fabrika, S., et al., 2010, {\it Astrophys. J.}, {\bf 709}, 1374

Lipunov V.M. \& Shakura, N.I., 1982, {\it Soviet Astron.}, {\bf 26}, 386

Margon, B., Ford, H.C., Grandy, S.A., Stone, R.P.S., 1979, {\it Astrophys. J.},
{\bf 233}, L63

Margon, B., 1984, {\it Annual Review Astron. Astrophys.}, {\bf 22}, 507

Marshall, H.L., Canizares, C.R., Schulz, N.S., 2002, {\it Astrophys. J.}, {\bf 564},
941

Massey, P., Meyer, M.R., 2001. Stellar Masses. In {\it Encyclopedia of Astron.
and Astrophysics}. Nature Publishing Group \& Institute of Physics Publishing.
Dirac House, Bristol, UK
(http://www.astro.caltech.edu/$\sim$george/ay20/eaa-stellarmasses.pdf)

Matthews, T. A. \& Sandage, A. R., 1963, {\it Astrophys. J.}, {\bf 138}, 30

Mironov, A.V., 2008, {\it Osnovy astrofotometrii} ({\it The Basics of Stellar
Photometry}). FizMatLit. Moscow. P. 207

Moro, D. \& Munari, U., 2000, {\it Astron. \& Astrophys. Suppl. Series},
{\bf 147}, 361

Murdin, P., Clark, D.H., Martin, P.G., 1980, {\it Monthly Not. Roy. Astron. Soc.},
{\bf 193}, 135

Nandy, K., Morgan, D.H., Willis, A.J., et al., 1981, {\it Monthly Not. Roy.
Astron. Soc.}, {\bf 196}, 955

Newsom, G.H. \& Collins II, G.W., 1981, {\it Astron. J.}, {\bf 86}, 1250

Newsom, G.H. \& Collins II, G.W., 1986, {\it Astron. J.}, {\bf 91}, 118

Paragi, Z., Vermeulen, R.C., Fejes, I., et al., 1999,
{\it Astron. \& Astrophys.}, {\bf 348}, 910

Prevot, M.L., Lequeux, J., Maurice E., et al., 1984, {\it Astron.
\& Astrophys.}, {\bf 132}, 389

Romney, J.D., Schilizzi, R.T., Fejes, I., \& Spencer, R.E., 1987, {\it Astrophys.
J.}, {\bf 321}, 822

Savage, B.D. \& Mathys, J.S., 1979, {\it Annual Review Astron. Astrophys.}, {\bf 17},
73

Schaller, G., Schaerer, D., Meynet, G., Maeder, A., 1992, {\it Astron.
\& Astrophys. Suppl. Series}, {\bf 96}, 269

Schild, R.E., 1977, {\it Astrophys. J.}, {\bf 82}, 337

Seaton, M.J., 1979, {\it Monthly Not. Roy. Astron. Soc.}, {\bf 187}, 73P

Spencer, R.E., 1984, {\it Monthly Not. Roy. Astron. Soc.}, {\bf 209}, 869

Stephenson, C.B. \& Sanduleak, N., 1977, {\it Astrophys. J. Suppl.}, {\bf 33}, 459

Straizys, V. {\it Multicolor Stellar Photometry}. Mokslas, Vilnius. 1977

Straizys, V. \& Kuriliene, G., 1975, {\it Bull. Vilnius Obs.}, {\bf 42}, 16

Thorne, K.S. \& Zytkow, A.N., 1975, {\it Astrophys. J.}, {\bf 199}, L19

Thorne, K.S. \& Zytkow, A.N., 1977, {\it Astrophys. J.}, {\bf 212}, 832

Vermeulen, R.C., Murdin, P.G., van den Heuvel, E.P.J., et al., 1993a,
{\it Astron. \& Astrophys.}, {\bf 270}, 204

Vermeulen, R.C., Schilizzi, R.T., Spencer, R.E., et al., 1993b,
{\it Astron. \& Astrophys.}, {\bf 270}, 177

Wagner, R.M., Newsom, G.H., Foltz, C.B., Byard, P.L., 1981, {\it Astron. J.},
{\bf 86}, 1671

Wagner, R.M., 1986, {\it Astrophys. J.}, {\bf 308}, 152

Yuan, W., Kawai, N., Brinkmann, W., \& Matsuoka, M., 1995,
{\it Astron. \& Astrophys.}, {\bf 297}, 451
\endreferences

\newpage
\begin{table}
  \caption{Mid-eclipse times}
  \smallskip
  \begin{center}
\begin{tabular}{lllcllllc}
\hline
\\
JD hel.  &$\sigma$&$\psi$& Source &\ \ \ \ \ \ \ \ \ \ \ \ \ \
  &JD hel.     &$\sigma$&$\psi$&Source\\
24...    &  day    &      &        &  &24...       &  day      &      &\\
\\
\hline
\\
44019.10  & 0.05 &  0.156& G      &  &   47394.310&  0.03 &  0.955& G \\
44332.73  & 0.05 &  0.088& G      &  &   47420.32 &  0.04 &  0.114& G \\
44463.85  & 0.04 &  0.896& G      &  &   48035.06 &  0.05 &  0.903& G \\
44476.93  & 0.03 &  0.975& G      &  &   48061.44 &  0.04 &  0.066& G \\
44489.97  & 0.05 &  0.058& G      &  &   49840.52 &  0.04 &  0.030& G \\
44790.89  & 0.06 &  0.910& G      &  &   49984.49 &  0.05 &  0.914& F \\
44816.87  & 0.05 &  0.073& G      &  &   50965.88 &  0.05 &       & G \\
45275.10  & 0.05 &  0.895& G      &  &   50978.61 &  0.10 &       & G \\
45928.95  & 0.04 &  0.925& G      &  &   51763.84 &  0.05 &       & G \\
45942.15  & 0.02 &  0.006& G      &  &   51776.82 &  0.05 &       & G \\
45955.18  & 0.10 &  0.085& G      &  &   52758.00 &  0.05 &  0.009& Ir\\
45968.46  & 0.10 &  0.167& G      &  &   52770.863&  0.05 &  0.088& Ch\\
46282.30  & 0.03 &  0.102& G      &  &   52914.77 &  0.08 &  0.975& FG\\
46583.05  & 0.10 &  0.955& G      &  &   53582.14 &  0.02 &  0.088& G \\
46596.10  & 0.04 &  0.037& G      &  &   53595.23 &  0.04 &  0.168& G \\
46936.3047& 0.013&  0.133& GI     &  &   54079.06 &  0.10 &  0.150& G:\\
46936.34  & 0.04 &  0.133& G      &  &   54380.33 &  0.02 &  0.006& K \\
47093.250 & 0.03 &  0.099& G      &  &            &       &       & \\
\\
\hline
\\
\multicolumn{9}{l}{\footnotesize
G: from the author's data collection; GI: from Ginga X-ray
observations; F: S.N. Fabrika;}\\
\multicolumn{9}{l}{\footnotesize
Ir: T.I. Irsmambetova; Ch:
Cherepashchuk et al. (2005); FG: S.N. Fabrika \& V.P. Goranskij; K: Kubota et al. (2010).}\\
\end{tabular}
\end{center}
\end{table}

\begin{table}
  \caption{Flux densities of a zero-magnitude A0V star in the units of
  $erg\cdot cm^{-2}\cdot s^{-1}\cdot A^{-1}$
}
  \smallskip
  \begin{center}
\begin{tabular}{cccccc}
\hline
\\
Band  &$\lambda_0$&$log_{10} \lambda_0$& Strayzys, & Moro \& Munari  & This paper\\
     &   \AA     &     \AA       &   (1977)    &   (2000)          &          \\
\\
\hline\\
U    &  3642     &   3.561       &$4.22\cdot10^{-9}$&$3.98\cdot10^{-9}$ &$4.024\cdot10^{-9}$ \\
B    &  4417     &   3.645       &$6.40\cdot10^{-9}$&$6.95\cdot10^{-9}$ &$6.426\cdot10^{-9}$ \\
V    &  5505     &   3.741       &$3.75\cdot10^{-9}$&$3.63\cdot10^{-9}$ &$3.713\cdot10^{-9}$ \\
R$_C$&  6469     &   3.811       &      -           &$2.254\cdot10^{-9}$&$2.300\cdot10^{-9}$\\
I$_C$&  7886     &   3.897       &      -           &$1.196\cdot10^{-9}$&$1.231\cdot10^{-9}$\\
\\
\hline
\end{tabular}
\end{center}
\end{table}

\begin{table}
  \caption{Effective wavelengths of light measured for SS 433 in the $UBVR_CI_C$
  filters and flux densities of a zero-magnitude A0V type star for this light
  in units of $erg\cdot cm^{-2}\cdot s^{-1}\cdot A^{-1}$
}
  \smallskip
  \begin{center}
\begin{tabular}{cccc}
\hline
\\
Mag  &$\lambda_{eff}$&$log_{10} \lambda_{eff}$& $f_\lambda$ \\
     &   \AA     &     \AA       &             \\
\\
\hline\\
U    &  3722     &   3.571       &$4.583\cdot10^{-9}$ \\
B    &  4623     &   3.665       &$5.876\cdot10^{-9}$ \\
V    &  5652     &   3.752       &$3.455\cdot10^{-9}$ \\
R$_C$&  6607     &   3.820       &$2.153\cdot10^{-9}$\\
I$_C$&  7891     &   3.897       &$1.228\cdot10^{-9}$\\
\\
\hline
\end{tabular}
\end{center}
\end{table}

\begin{table}
  \caption{Contribution of emission-line radiation to $UBV$ filters
for the eclipse of 2007 October 6 (percents)}
  \smallskip
  \begin{center}
\begin{tabular}{lcccc}
\hline
\\
Telescope  &$U^*$&$B$&$V$& $\phi$ \\
\\
\hline\\
Subaru    &  13.6     &   15.2 & 7.6  & 0.984\\
BTA       &   -       &   18.2 & 5.1  & 0.018\\
\\
\hline
\multicolumn{5}{l}{\footnotesize
* Balmer continuum radiation did not taken into account.}
\end{tabular}
\end{center}
\end{table}

\begin{table}
  \caption{Theoretical mass - luminosity relation for A stars
  based on computations in Schaller et al. (1995)}
  \smallskip
  \begin{center}
\begin{tabular}{ccccc}
\hline
\\
Initial & Current & $log M_A$ & $M_{bol}$ & $log L/L_\odot$\\
mass    & mass    &\\
\\
\hline\\
4   & 4.0    & 0.602 & -2.10  & 2.740\\
5   & 5.0    & 0.699 & -3.00  & 3.100\\
7   & 7.0    & 0.845 & -4.35  & 3.640\\
9   & 8.99   & 0.956 & -5.24  & 3.996\\
12  & 11.90  & 1.076 & -6.22  & 4.338\\
15  & 14.70  & 1.167 & -7.02  & 4.708\\
20  & 19.06  & 1.280 & -8.18  & 5.172\\
25  & 23.62  & 1.373 & -8.70  & 5.380\\
40  & 36.06  & 1.557 & -9.50  & 5.700\\
\\
\hline
\end{tabular}
\end{center}
\end{table}

\end{document}